\documentclass[prb,aps,twocolumn,floatfix,showpacs]{revtex4}

\usepackage{amssymb}
\usepackage{amsmath}
\usepackage{graphicx}
\usepackage{amssymb}

\newcommand{\unit}[1]{\rm{#1}}

\begin{document}

\preprint{}
\title[ShortTitle]{Aharonov-Bohm phase as quantum gate in two-electron charge qubits}
\author{A.~Weichselbaum}
\author{S.~E.~Ulloa}
\affiliation{Department of Physics and Astronomy, Nanoscale and
Quantum Phenomena Institute, Ohio University, Athens, Ohio 45701}



\begin{abstract}
We analyze the singlet-triplet splitting on a planar array of
quantum dots coupled capacitively to a set of external voltage
gates. The system is modelled using an extended Hubbard
Hamiltonian keeping two excess electrons on the array. The voltage
dependence of the low-energy singlet and triplet states is
analyzed using the Feshbach formalism. The formation of a well
decoupled two-level system in the ground state is shown to rely on
the fact of having two particles in the system. Coherent operation
of the array is studied with respect to single quantum bit
operations. One quantum gate is implemented via voltage controls,
while for the necessary second quantum gate, a uniform external
magnetic field is introduced. The Aharonov-Bohm phases on the
closed loop tunnel connections in the array are used to
effectively suppress the tunneling, despite a constant tunneling
amplitude in the structure. This allows one to completely stall
the qubit in any arbitrary quantum superposition, providing full
control of this interesting quantum system.
\end{abstract}

\volumeyear{year}
\volumenumber{number}
\issuenumber{number}
\eid{identifier}
\date[Date: ]{Mar 02, 2004}
\maketitle

\section{Introduction}

Semiconductor quantum dot systems have been studied extensively in
recent years. These quantum dots (qudots) typically contain from a
few to a few hundreds of extra-electrons and are externally
controlled by voltage gates. Their behavior at low bias voltages
can be well understood by looking at the energetically topmost
electrons in the qudot. \cite{Tarucha00} For an even number of
electrons, they may pair up such that the total spin is zero and
the topmost two electrons essentially form a singlet, a system
that has been proposed as a possible source for entangled
electrons. \cite{Saraga03} However, under special circumstances,
the ground state may actually be a triplet even without the
presence of an external magnetic field. \cite{Fuhrer03} A
straightforward explanation for this behavior can be given in
terms of the exchange contribution to the energy from the two
topmost electrons for the case of nearly degenerate single
particle levels. \cite{Tarucha00} For an odd total number of
electrons on a single qudot, the overall spin is typically found
to be $1/2$.

In this paper we consider an ensemble of qudots, specifically a
$2\times 2$ array of interconnected and interacting qudots similar
to the typical cellular automata unit cell format. \cite{Toth01,
Amlani99, Lent00, Lieber02, Rojas04} The interaction is modelled
within a capacitance matrix formalism. In the weak tunneling
regime, the direct exchange energy from the two-body interaction
between electrons on different qudots is negligible. Residence of
these two electrons on the same qudot (double occupancy) is
energetically unfavorable due to the comparatively large Coulomb
charging energies. The $2\times 2$ array is considered as a
particular implementation of a charge quantum bit (charge qubit).
As shown in [\onlinecite{Wb04}], it is essential for the single
qubit operations of such a charge qubit to have a non-local
potential, tuneable tunneling amplitudes or an external magnetic
field which provides a complex phase to the wavefunction. Since
the direct exchange contribution with charges on different dots is
negligible, there are no non-local potential effects here.
Further, the tunneling amplitude is considered constant, fixed by
the specific realization of the solid state qudot array which
typically uses oxide barriers between qudots, and are then
basically unaltered by potential gates. In this context of a fixed
geometry, the only way to implement full single qubit operations
is using an external magnetic field which can be controlled at
will. \cite{Wb04} It is further considered that the field is
uniform within the area of the qudot array as is likely the case
in normal implementations.

Due to the geometrical symmetry of the qudot array, the ground
state turns out to be exactly degenerate for the triplet states
when no magnetic field and no gate voltages are applied. However,
the spectrum exhibits a gap for the singlet states which is
related to the distinct symmetry under particle exchange for the
singlet and triplet states. Therefore, particle exchange does play
an essential role, yet it is a higher order effect, similar to
superexchange \cite{Jeff96}, as the exchange of two particles on
the qudot array takes more than one tunneling step and explores
virtual higher energy states.

The properties of the singlet and triplet states are determined by
two remaining parameters, the asymmetrically applied gate voltage
$V_{g}$, which breaks the $90$ degree symmetry, and the external
magnetic field $B$ perpendicular to the array. We show that
careful control of the magnetic flux through the system allows one
to rotate the qubit Bloch vector about an axis orthogonal to the
one provided by $V_{g}$. These two parameters then, provide full
quantum manipulation of the charge qubits.

The analysis of our numerical results follows the Feshbach
formalism (App.~\ref{APP.Feshbach}) which provides an effective
Hamiltonian of the two relevant dimensions of the qubit subspace,
well decoupled energetically from higher lying states. This
approach provides insights on the nature of the system dependence
on fields, and explains the ability of the magnetic field to
complete the set of necessary single qubit operations.

\section{The Model System}

We examined a range of different geometries of planar arrays with
capacitively coupled quantum dots. The best decoupled quantum $2$
level systems (qu2LS) in the ground state were found to exhibit
high spatial symmetry, i.e. $C_{4v}$ symmetry, in agreement with
the requirement of spatially distinct wave functions needed to
participate in the qu2LS. \cite{Wb04} The $C_{4v}$ symmetry of the
$2\times 2$ array, for example, ensures a binary groundstate
system that for weak inter-dot tunneling is well decoupled
energetically from the remainder of the spectrum. Other geometries
are clearly possible, but we have chosen here one of the simplest
closed loops with $C_{4v}$ symmetry to demonstrate the main
concepts.

\begin{figure}[tbp]
\includegraphics*[angle=0,width=8.6cm]{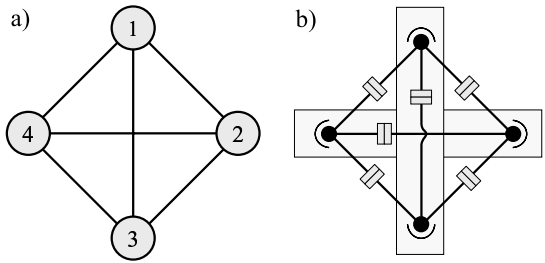}
\caption{Setup of $2\times 2$ array. (a) Arrangement of the four
islands with mutual tunnel connection indicated by black lines.
(b) Same as in (a) but shows explicitly tunneling junctions and
capacitive coupling including the two voltage gates acting along
the square diagonals.} \label{fig_2x2_setup}
\end{figure}

The model network under consideration is a $2\times 2$ array of
qudots with a single spatial state per site plus spin. The system
is sketched in Fig.~\ref{fig_2x2_setup}a. As explicitly outlined
in panel (b), tunneling is allowed between any pair of dots, where
every tunnel junction also carries capacitance. The parameters
which enter the model are: the dot-gate capacitance
($C_{g}=25\mathrm{aF}$), the dot-dot capacitances
($C_{dd}=25\mathrm{aF}$ for nearest neighbor dots and
$17\mathrm{aF}$ for dots connected through the diagonal of the
array), the dot self-capacitance ($C_{d0}=25\mathrm{aF}$) and the
dot-dot tunneling amplitude ($t=2\mathrm{\mu eV}$). The parameters
have been chosen such that the energy cost for double occupancy of
a dot (standard Hubbard $U$) is about $1\mathrm{meV}$, a commonly
used value for these systems, and corresponding to typical dot
dimensions of $100\mathrm{nm}$. Further, the tunneling $t$ must
not be chosen too large since otherwise the quality of the
$2$-level system in the groundstate as well as the well-defined
charge per dot are compromised, as we will see below.

The Hamiltonian used to describe this system is of the extended Hubbard type

\begin{eqnarray}
H &=&\sum_{i,\sigma }\varepsilon _{\sigma }\,c_{i\sigma }^{+}c_{i\sigma
}-\sum_{i,j,\sigma }t_{ij}^{\sigma }(c_{i\sigma }^{+}c_{j\sigma }+c_{j\sigma
}^{+}c_{i\sigma })+  \notag \\
&&\frac{1}{2}\sum_{i,j,\sigma ,\sigma ^{\prime }}V_{ij\,}c_{i\sigma
}^{+}c_{j\sigma ^{\prime }}^{+}c_{j\sigma ^{\prime }}c_{i\sigma
}+\sum_{i}V_{i\,}\hat{n}_{i}  \label{Hubbard_H}
\end{eqnarray}%
with $\hat{n}_{i}\equiv c_{i\uparrow }^{+}c_{i\uparrow
}+c_{i\downarrow }^{+}c_{i\downarrow }$ and $c_{i\sigma}^{+}$ the
typical creation operator for a particle at qudot $i$ with spin
$\sigma$. $\varepsilon _{(i)\sigma }$ refers to the local energy
of state $\sigma $ on the $i=\{1,\ldots ,n\}$ identical dots and
can be used to account for the Zeeman splitting of spins in an
external magnetic field. Throughout this paper, however, the
$\varepsilon _{\sigma }$ are simply set equal and zero. The
tunneling coefficients $t_{ij}^{\sigma }$ from dot $i$ to dot $j$
are considered independent of the spin orientation, thus
$t_{ij}^{\uparrow }=t_{ij}^{\downarrow }\equiv t_{ij}$.
Furthermore, the tunnel connections are considered to be the same
up to a phase, i.e. $\left\vert t_{ij}\right\vert \equiv
\left\vert t\right\vert$. The electrostatic energy in the last two
terms of Eq.~(\ref{Hubbard_H}), i.e. the coefficients $V_{ij}$ and
$V_{i}$, are derived from the total capacitance matrix of the
system which is approximated by the capacitor network indicated in
Fig.~\ref{fig_2x2_setup}b. \cite{Wb04}

An essential property of singlet and triplet states is the
effective separation of the spin degree of freedom from the
spatial wavefunction component since the total state is a product
of spatial and spin components. Here the (anti)symmetry of the
spatial wavefunction of (triplet) singlet states under particle
exchange is taken care of by imposing (anti)commutator relations
on the creation and annihilation operators $c_{i}^{+}$ and $c_{i}$
with the spin index dropped, yet keeping the constraint of a total
of two electrons.

\begin{figure}[tbp]
\includegraphics*[angle=0,width=8.6cm]{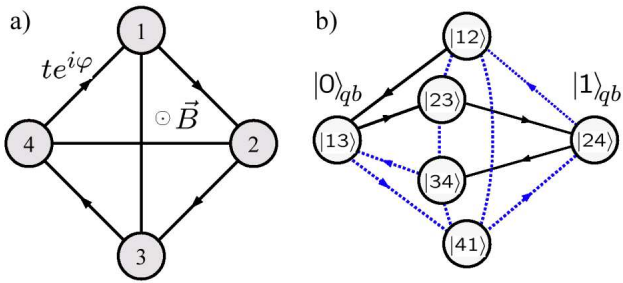}
\caption{$2\times 2$ qudot array. (a) Array with a perpendicular
magnetic field applied. (b) Schematic network of the Hilbert space
states with tunneling transitions between them indicated by
connecting lines (states of double occupancy are not included).
The arrows indicate the flow of complex phase acquired from a
magnetic field in the tunnelling coefficients $\left\vert
t\right\vert e^{i\protect \varphi }$. The dashed blue lines
indicate paths of particle exchange (see text). $\left\vert
0\right\rangle _{qb} \equiv \left\vert 13\right\rangle$ and
$\left\vert 1\right\rangle _{qb} \equiv \left\vert
24\right\rangle$ indicate the qubit states where $\left\vert
ij\right\rangle$ are the two electron states with one electron on
dot $i$ and the other on dot $j$.} \label{fig_2x2_Bhilbert}
\end{figure}

An external magnetic field perpendicular to the network of qudots
affects the system insofar as spatial propagation is associated
with the acquisition of a complex phase. Therefore the $t_{ij}$
become complex \cite{Peierls33}
\begin{equation}
t_{ij}=t_{ji}^{\ast }=\left\vert t_{ij}\right\vert e^{i\varphi
_{ij}}, \text{ with }\varphi _{ij}\equiv \frac{e}{\hbar
}\int\limits_{x_{i}}^{x_{j}}\vec{A} \cdot d\vec{\ell},
\label{phases_duetoB}
\end{equation}%
where $\vec{A}$ is the vector potential of the applied magnetic
field, $\vec{B}=\vec{\nabla}\times \vec{A}$. Using a symmetric
gauge, the acquisition of phase in the $2\times 2$ array of qudots
is indicated in Fig.~\ref{fig_2x2_Bhilbert}a. The phase
\textit{flows }clockwise on the outer connections while the
diagonal connections remain phaseless. The ring structure leads to
an Aharonov-Bohm phase (AB) for a single particle moving around
the ring. Note, however, that the second particle is essential for
the necessary ground state qu2LS needed in the qubit setup. For
lithographic setups on the scale of $200\mathrm{nm}$, the required
magnetic field for an AB phase cycle on the whole array is around
$100\mathrm{mT}$ and thus rather small. At these fields the local
wave functions in the individual quantum dots do not change much,
and the absolute value of the tunneling $\left\vert t\right\vert $
is considered constant.

In Fig.~\ref{fig_2x2_Bhilbert}b, the two-particle Hilbert space
and the allowed tunnel transitions are shown for the $2\times 2$
system, for simplicity ignoring states of double occupancy. The
blue dashed lines in Fig.~\ref{fig_2x2_Bhilbert}b are related to
particle exchange in the sense that the off-diagonal element in
the corresponding triplet Hamiltonian has an extra minus sign due
to its fermionic character, as it can also be seen directly from
the basis chosen. For example, the transition $\left\vert
12\right\rangle$ to $\left\vert 23\right\rangle$ can be thought of
as the two-step process of one hopping and one exchange,
$\left\vert 12\right\rangle \rightarrow \left\vert 32\right\rangle
\rightarrow -\left\vert 23\right\rangle$. Any path with an odd
number of dashed segments in the Hilbert space of
Fig.~\ref{fig_2x2_Bhilbert}b has an extra minus sign associated
with it. Notice also, that if the path in the Hilbert space
network of Fig.~\ref{fig_2x2_Bhilbert}b is closed, then an odd
number of dashed segments refers to an effective exchange of the
two electrons. More on this later.

\section{Analysis}

\begin{figure}[tbp]
\includegraphics*[angle=0,width=8.6cm]{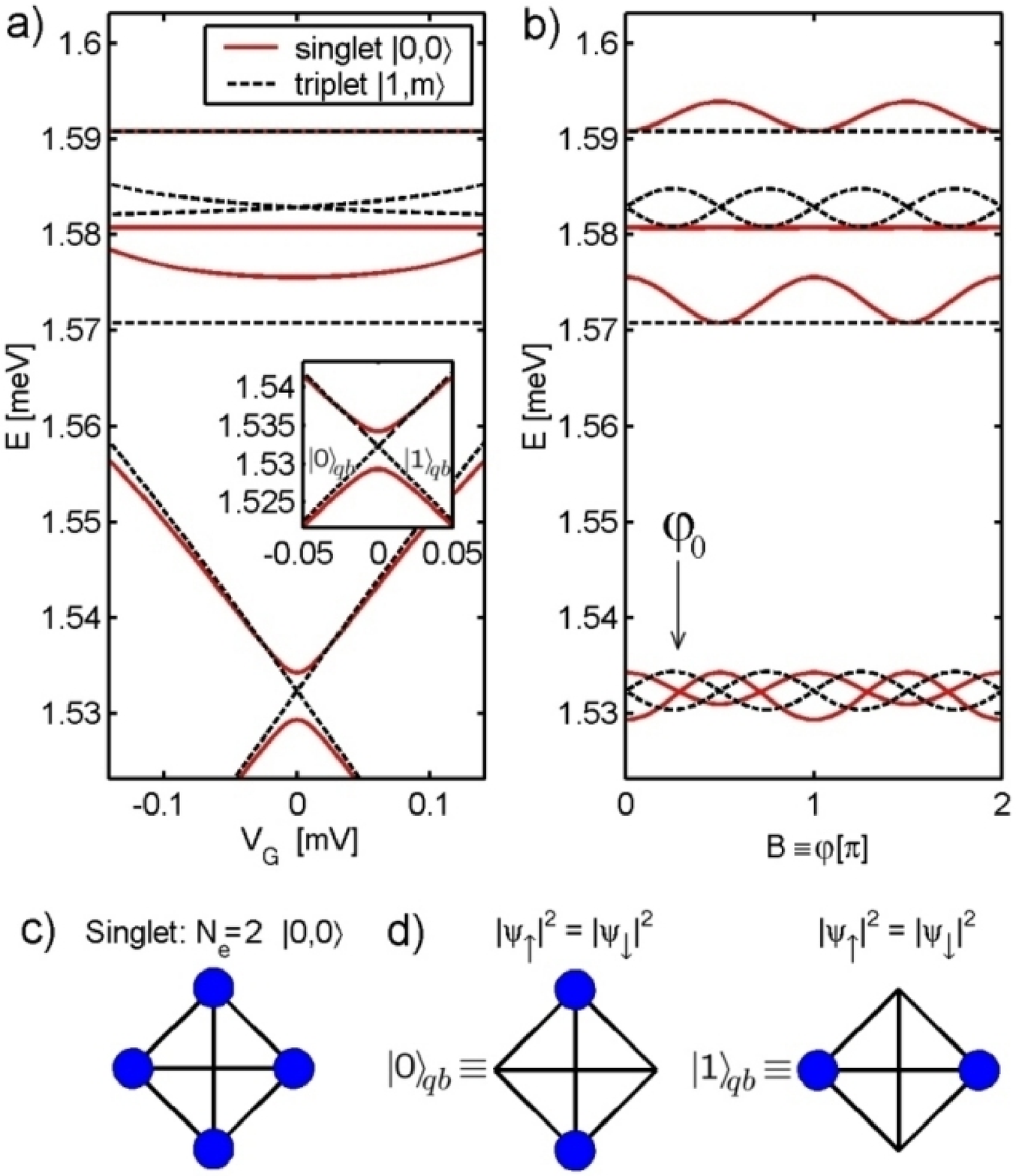}
\caption{Energy spectra for the qu2LS of the $2\times 2$ qudot
system together with a few higher lying states for singlet and
triplet states ($\left\vert S,~S_{z}\right\rangle =\left\vert
0,0\right\rangle $ and $\left\vert S,~S_{z}\right\rangle
=\left\vert 1,m=-1,0,1\right\rangle $ respectively). (a) Energy
spectrum vs. asymmetrically applied gate voltage, $V_{G} \equiv
V_{g_{1}}=-V_{g_{2}}$. The doubly occupied states lie about $1
\mathrm{meV}$ higher in energy (outside figure) and therefore have
negligible influence. The inset shows a closeup of the
(anti)crossing in the qu2LS. (b) Energy spectrum vs. uniform
external magnetic field perpendicular to the array expressed
through the phase in $t=\left\vert t\right\vert
e^{i\protect\varphi }$. The initial singlet anticrossing at
$\protect \varphi =0$ is completely closed for $\protect \varphi
=\protect \varphi _{0} =0.286~\protect\pi $, indicated by the
arrow in panel (b), while at the same time the triplet levels show
a pronounced anticrossing. (c) Singlet ground state probability
distribution over the $2\times 2$ array. This state is a symmetric
combination of the basis states shown in panel (d): Probability
distribution of the basis states of the singlet qu2LS labelled
$\left\vert 0\right\rangle _{qb}$ and $\left\vert 1\right\rangle
_{qb}$ with equal probability to find spin up or spin down,
$\left\vert \protect\psi _{\uparrow }\right\vert ^{2}=\left\vert
\protect\psi _{\downarrow }\right\vert ^{2}$.}
\label{fig_2x2_spectrum}
\end{figure}

Using the Hamiltonian in Eq.~(\ref{Hubbard_H}), the eigenspectrum
for the $2\times 2$ array is shown in Fig.~\ref{fig_2x2_spectrum}a
as function of an asymmetric gate voltage drag $V_{g}$, with no
magnetic field applied ($B=0$). For small $V_{g}$, the singlet
(triplet) low-energy level set of interest is well separated from
the remaining singlet (triplet) spectrum. In contrast to the
singlet set, however, which has an anticrossing at $V_{g}=0$, the
triplet set is degenerate there. Panel (b) shows the eigenspectrum
as function of the magnetic field when there are no gate voltages
applied ($V_{g}=0$). The tunneling has been chosen relatively
large ($\left\vert t\right\vert =5\mathrm{\mu eV}$) such that the
energy splitting $\delta $ due to the tunneling in the low-energy
singlet set (the qu2LS we will focus on) reaches about $1/10$ of
the distance to the nearest higher lying states, $\Delta $. This
is still a good qubit configuration, as the coherent state
manipulation in the qu2LS can be performed without significant
admixture of the higher lying states. However, the gates must be
switched smoothly enough for the evolution to be adiabatic with
respect to the higher lying states, as will be seen later from the
numerical analysis.

The energy spectrum in Fig.~\ref{fig_2x2_spectrum}b is periodic in
the magnetic field in the usual AB sense. Since $\varphi$ relates
to one quarter of the phase on the entire outer loop, this means
that with every $\Delta \varphi =2\pi/4$ one additional flux
quantum enters or leaves the cross-sectional area of the array.
This is seen for example in the splitting of the singlet which
opens and closes with a period $\Delta \varphi =2\pi/4$. The exact
period of the system, however, is $\Delta \varphi =\pi$. Note,
that this is not because of the usual $t\rightarrow -t$ symmetry
which does not hold here because of the diagonal cross link in the
array shown in Fig.~\ref{fig_2x2_Bhilbert}a. Instead, it can be
related to changing the sign in both basis states of the qu2LS.
This is easily seen for $\varphi =\pi$ from
Fig.~\ref{fig_2x2_Bhilbert}b by considering $\left\vert
0\right\rangle _{qb} \equiv \left\vert 13\right\rangle =
-\left\vert 31\right\rangle$ and dropping all the arrows shown.

Now the essential effect of the magnetic field is that it allows
one to close the gap in the singlet qu2LS while at the same time
it opens a gap in the triplet qu2LS (at fixed $S_z$). The smallest
magnetic field where this happens is at $\varphi \equiv \varphi
_{0}=0.286~\pi $, indicated by the arrow in
Fig.~\ref{fig_2x2_spectrum}b. The important consequence of tuning
the magnetic field to $\varphi \rightarrow \varphi _{0}$ is that
the charge qubit can be held \textit{frozen} in its state by also
having $V_{g}=0$ (see below).

Figure \ref{fig_2x2_spectrum}c and d show the singlet ground state
configuration and its two \textit{spatially distinct }basis
states, respectively. Note that in order to have (close to)
degenerate groundstate qu2LS, there must exist a basis
representation that is spatially complementary \cite{Wb04}, in
agreement with what is shown in panel (d).

In the context of quantum computation, any linear superposition of
the two states forming the qubit must be possible.
\cite{Nielsen00} The problem is conveniently mapped into a pseudo
spin Hamiltonian with its equivalent $2$-level system. The general
qubit state $\left\vert \psi \right\rangle =\left(
c_{1},c_{2}\right) $ when written as a density matrix $\rho
=\left\vert \psi \right\rangle \left\langle \psi \right\vert
\equiv \frac{1}{2}\left( 1+\vec{r }_{b}\vec{\sigma}\right) $
defines the $3D$ Bloch vector $\vec{r}_{b}$ which then is used as
the representation of the qubit state. \cite{Nielsen00} Within
this frame, an arbitrary qubit operation translates into a
rotation of the Bloch vector (pseudo spin) anywhere in its $3$D
domain. Two distinct rotations are sufficient to do so. The
rotations in pseudo spin language are then generated by the Pauli
matrices $\sigma _{\{x,y,z\}}$ and the requirement of two distinct
rotations translates into two distinct quantum gates (qugates)
that can be built from the Pauli matrices.

For the charge qubit encoded in the $2\times 2$ array with the
basis states as shown in Fig.~\ref{fig_2x2_spectrum}d, the
physical quantum gates are now as follows: the asymmetrically
applied gate voltage (see Fig.~\ref{fig_2x2_setup}b) only drags
apart the potentials of the two qubit basis states in
Fig.~\ref{fig_2x2_spectrum}d and thus represents the $\sigma
_{z}$-gate, also referred to as $V$-gate. On the other hand, the
singlet gap or anticrossing can be related to a real off-diagonal
element in the $2$D pseudospin Hamiltonian which can be tuned down
to exactly zero by a magnetic field as shown above. Thus this is
referred to as the $\sigma _{x}$-gate or $B$-gate. Together, the
two physical qugates introduced can be utilized to generate
arbitrary rotations of the Bloch vector and thus to construct
arbitrary qugates for the qu2LS. Also, since both of the qugates
can be turned off completely by setting $V_{g}=0$ and turning on a
specified magnetic field, this allows to freeze the qu2LS in any
arbitrary state at any time.

\subsection{Splitting due to exchange energy\label{SubSec.XSplitting}}

We want to analyze now the reasons for the closing and opening of
the gap in the qu2LS, as a way to provide us with insights into
the nature of these states. The Hamiltonian in
Eq.~(\ref{Hubbard_H}) is written out explicitly as a
six-dimensional matrix in Eq.~(\ref{2x2_H}) for two electrons on
the $2\times 2$ array where, for simplicity, states of double
occupancy are neglected. %
\begin{equation}
H=\left(
\begin{array}{c|cc|cccc}
& \left\vert 13\right\rangle & \left\vert 24\right\rangle & \left\vert
12\right\rangle & \left\vert 23\right\rangle & \left\vert 34\right\rangle &
\left\vert 41\right\rangle \\ \hline
\left\vert 13\right\rangle & \varepsilon _{1} & 0 & -t^{\ast } & -t & \pm
t^{\ast } & -t \\
\left\vert 24\right\rangle &  & \varepsilon _{1} & \pm t & -t^{\ast } & -t &
\pm t^{\ast } \\ \hline
\left\vert 12\right\rangle &  &  & \varepsilon _{2} & \pm \left\vert
t\right\vert & 0 & \pm t \\
\left\vert 23\right\rangle &  &  &  & \varepsilon _{2} & \pm \left\vert
t\right\vert & 0 \\
\left\vert 34\right\rangle &  &  &  &  & \varepsilon _{2} & \pm \left\vert
t\right\vert \\
\left\vert 41\right\rangle & c.c. &  &  &  &  & \varepsilon _{2}%
\end{array}%
\right)  \label{2x2_H}
\end{equation}%
The sign in $-$($+$)$t$ refers to the singlet (triplet) spatial
Hamiltonian, respectively, with the complex tunneling $t$ as in
Eq.~(\ref{phases_duetoB}). The top row and the left-most column of
Eq.~(\ref{2x2_H}) indicate the two-particle basis states chosen as
shown in Fig.~\ref{fig_2x2_Bhilbert}b. With $t=0$, the qu2LS
grouped in the upper left $2\times 2$ block of the $H$ matrix is
degenerate and all intermediate states are split off by the same
energy $\Delta _{0}\equiv \varepsilon _{2}-\varepsilon _{1}$ due
to symmetry. Here, $\varepsilon _{1}$ and $\varepsilon _{2}$ are
the overall diagonal contributions arising from the Coulomb
interaction at $V_g =0$.

The Hamiltonian in Eq.~(\ref{2x2_H}) can be diagonalized
analytically. However, a perturbative approach provides explicit
insights on the reasons for the splitting due to exchange of
otherwise degenerate states. Furthermore, the setup proves
sufficiently simple to allow the complex sum to all orders over
all possible histories in Hilbert space within the Feshbach
formalism. The result consistently agrees with the analytical
solution to the problem.

In order to obtain an estimate for the splitting in the ground
state set, the Feshbach formalism provides an effective
Hamiltonian for the qu2LS ($\mathcal{P}$ space) coupled to higher
lying states ($\mathcal{Q}$ space, see App.~\ref{APP.Feshbach}).
For convenience, the qu2LS basis is written as $\mathcal{P}\equiv
\left\{ \left\vert 13\right\rangle ,\left\vert 24\right\rangle
\right\} \equiv \left\{ \left\vert 0\right\rangle _{qb},\left\vert
1\right\rangle _{qb}\right\} \equiv \left\{ 0, 1\right\} _{qb}$.
The remaining intermediate states in Fig.~\ref{fig_2x2_Bhilbert}b
($\mathcal{Q}$ space) form a ring topology where every node is
linked to the $\left\vert 0\right\rangle _{qb}$ and $\left\vert
1\right\rangle _{qb}$ states. Note that there is no direct
transition from $\left\vert 0\right\rangle _{qb}$ to $\left\vert
1\right\rangle _{qb}$, but one has to proceed through at least one
of the intermediate states with an energy cost of $\Delta
_{0}\equiv \varepsilon _{2}-\varepsilon _{1}$.

The effective $2D$ Hamiltonian for the qu2LS in the absence of a
magnetic field is now constructed as follows: the matrix element
$\left( H_{\mathit{eff}}\right) _{ij}$ with $i,j=\left\{
0,1\right\} _{qb}$ is the sum over all possible paths that start
in state $i$, immediately proceed to intermediate states
$\mathcal{Q}$, and only in the final step come back to state $j$.
The number of possible paths constructed in this manner for a
total of $n$ steps is $S_{n}\left( i,j\right) \equiv 2^{n}$ with
$n\geq 2$ since there are $4$ possibilities to go from $\left\vert
i\right\rangle $ to the intermediate states, then $2$
possibilities of which way to go in the ring at each of the $n-2$
intermediate steps, and $1$ choice left to finally leave the ring
and go to state $\left\vert j\right\rangle $. Furthermore, for
$n>2$, exactly half of these paths include particle exchange, i.e.
have an odd number of dashed lines in
Fig.~\ref{fig_2x2_Bhilbert}b, and thus for the case of the triplet
states cancel each other to zero. The underlying reason for this
is the spatial $C_{4v}$ symmetry of the $2\times 2$ setup which
has two mirror symmetry planes perpendicular to the array.
Therefore, for every path starting from $\left\{ 0,1\right\}
_{qb}$ and ending in $\left\{ 1,0\right\} _{qb}$ has a mirrored
counterpart where the particles are exchanged in the final state
as compared to the first path.

The case $n=2$ needs separate consideration. For the diagonal
elements in $H_{\mathit{eff}}$ the same step is taken twice, back
and forth, and thus the relative sign in $t$ does not matter.
Therefore the triplet states have an $n=2$ contribution in the
diagonal. Putting all these pieces together, yields the matrix
elements for the effective Hamiltonian which, for example, in case
of the singlet states are
\begin{eqnarray}
&&\left\langle i\right\vert H_{\mathit{eff}}\left\vert j\right\rangle
_{S}=\varepsilon _{1}\delta _{ij}+\frac{\left( -2t\right) ^{2}}{\omega
-\varepsilon _{2}}\sum_{m=0}^{\infty }\left( \frac{-2t}{\omega -\varepsilon
_{2}}\right) ^{m}  \notag \\
&=&\varepsilon _{1}\delta _{ij}+\frac{\left( -2t\right)
^{2}}{\omega -\varepsilon _{2}+2t}\equiv \varepsilon _{1}\delta
_{ij}+\Sigma _{ij}^{S}\left( \omega \right),  \label{2x2_Heff_ij}
\end{eqnarray}%
and thus%
\begin{subequations}
\begin{eqnarray}
H_{\mathit{eff}}^{singlet}\left( \omega \right) &\equiv &%
\begin{pmatrix}
\varepsilon _{1} & 0 \\
0 & \varepsilon _{1}%
\end{pmatrix}%
+\frac{4t^{2}}{\omega -\varepsilon _{2}+2t}%
\begin{pmatrix}
1 & 1 \\
1 & 1%
\end{pmatrix}
\label{2x2_HeffS_0} \\
H_{\mathit{eff}}^{triplet}\left( \omega \right) &\equiv &%
\begin{pmatrix}
\varepsilon _{1} & 0 \\
0 & \varepsilon _{1}%
\end{pmatrix}%
+\frac{4t^{2}}{\omega -\varepsilon _{2}}%
\begin{pmatrix}
1 & 0 \\
0 & 1%
\end{pmatrix},
\label{2x2_HeffT_0}
\end{eqnarray}
\label{2x2_Heff_0}
\end{subequations} %
where for the triplet state only the paths with $n=2$ contribute.
The eigenstates for the qu2LS are now obtained from the nonlinear
eigensystem $H_{\mathit{eff}}\left( \omega \right) \left\vert \psi
\right\rangle =\omega \left\vert \psi \right\rangle $ where
$\left\vert \psi \right\rangle $ is restricted to the $2D$ ground
space. $H_{\mathit{eff}}^{triplet}$ is still diagonal, and
therefore the triplet states \textit{do not mix} with each other,
but are just shifted lower by $\frac{\Delta _{0}}{2}-\sqrt{\left(
\frac{\Delta _{0}}{2}\right) ^{2}+4t^{2}}=-\frac{4t^{2}}{\Delta
_{0}}$ $+\mathcal{O} \left( t^{3}\right) $, with $\Delta _0 =
\varepsilon _2 - \varepsilon _1$. The singlet states, however,
rearrange to symmetric and antisymmetric combinations of the
original basis. One of the eigenstates remains at the original
eigenenergy $\omega =\varepsilon _{1}$, while the second one is
lowered by $\delta $ given as
\begin{equation}
\delta \equiv \left( \frac{\Delta _{0}}{2}-t\right) -\sqrt{\left( \frac{%
\Delta _{0}}{2}-t\right) ^{2}+8t^{2}}=-\frac{8t^{2}}{\Delta _{0}}+\mathcal{O}%
\left( t^{3}\right)  \label{splitting_B=0}
\end{equation}%
and thus forms the ground state of the system for finite $t$. The
analytical solutions are consistent with the numerical
diagonalization of the Hamiltonian in Eq.~(\ref{2x2_H}).

\subsection{Effect of a magnetic field\label{SubSec.Bsplitting}}

We apply again the Feshbach formalism to the system in
Fig.~\ref{fig_2x2_Bhilbert}b, but now including the complex phases
indicated by the arrows in that figure. Note that the phases due
to the magnetic field affect only the first and last step in each
path, while intermediate state transitions remain unaltered by the
presence of the magnetic field since they correspond to
transitions through the diagonal of the array in
Fig.~\ref{fig_2x2_Bhilbert}a (there are no arrows on the
transitions between intermediate states in
Fig.~\ref{fig_2x2_Bhilbert}b).

\begin{figure}[tb]
\includegraphics*[angle=0,width=8.6cm]{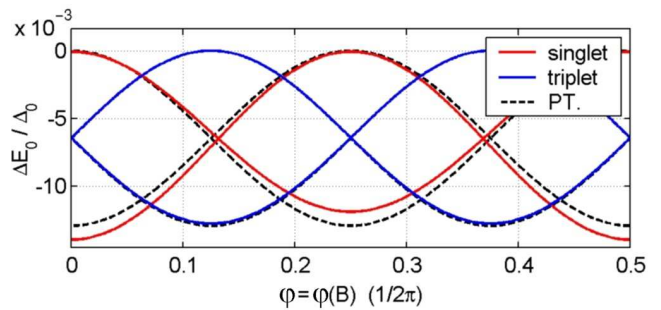}
\caption{Energy splitting in the qu2LS for the system of
Eq.~\ref{2x2_H} in dependence of the magnetic field for singlet
(red) and triplet states (blue). The splitting is shown in units
of $\Delta _{0}=\protect\varepsilon _{2}- \protect\varepsilon _{1}
$, namely the separation of the qu2LS from the remaining Hilbert
space. The dashed black line is the result of the lowest order
Feshbach analysis, Eq.~(\ref{2x2_Self_B}).}
\label{fig_2x2_B_dE_anal}
\end{figure}

The paths can be summed up similarly. For simplicity, however, we
give only the contribution to lowest order in $t$ for the
effective Hamiltonian. The denominator $\omega -\varepsilon _{2}$
is then replaced by $\varepsilon _{1}-\varepsilon _{2}=-\Delta
_{0}$, and so the lowest order contribution to the self energy
term is
\begin{subequations}
\begin{eqnarray}
\hspace{-5mm}\Sigma _{\mathit{eff}}^{singlet}\left( \omega ,\varphi \right)
\hspace{-1mm}&=&\hspace{-1mm} -\frac{4t^{2}}{\Delta
_{0}}%
\begin{pmatrix}
1 & \cos 2\varphi \\
\cos 2\varphi & 1%
\end{pmatrix}%
+\mathcal{O}\left( t^{3}\right)  \label{2x2_HeffS_B} \\
\hspace{-5mm}\Sigma _{\mathit{eff}}^{triplet}\left( \omega ,\varphi \right)
\hspace{-1mm}&=&\hspace{-1mm}-\frac{4t^{2}}{\Delta_{0}}%
\left(
\hspace{-3mm}
\begin{array}{cc}
1 & \hspace{-3mm}i\sin 2\varphi \\
-i\sin 2\varphi & \hspace{-3mm}1%
\end{array}
\hspace{-1.5mm}
\right) %
+\mathcal{O}\left( t^{3}\right). \label{2x2_HeffT_B}
\end{eqnarray}%
\label{2x2_Self_B}
\end{subequations}
Comparing this with the last terms in Eqs.~(\ref{2x2_Heff_0}), the
effect of the external magnetic field is obvious. With increasing
$\varphi $ the singlet splitting can be reduced down to zero,
while simultaneously a comparable gap opens in the triplet states,
in agreement with the numerical data for the full $2\times 2$
system in Fig.~\ref{fig_2x2_spectrum}b. The effect of the magnetic
field on the singlet (triplet) states is that of a $\sigma _{x}$
($\sigma _{y}$) gate, and thus clearly provides the necessary
second quantum gate for single qubit operation.
Figure~\ref{fig_2x2_B_dE_anal} compares the exact numerical
results of the Hamiltonian in Eq.~(\ref{2x2_H}) with above lowest
order perturbative approach for the splitting in the qu2LS. The
lowest order contribution in the Feshbach formalism already
provides an excellent approximation.

\subsection{Numerical qubit dynamic}

\begin{figure}[tb]
\includegraphics*[angle=0,width=8.6cm]{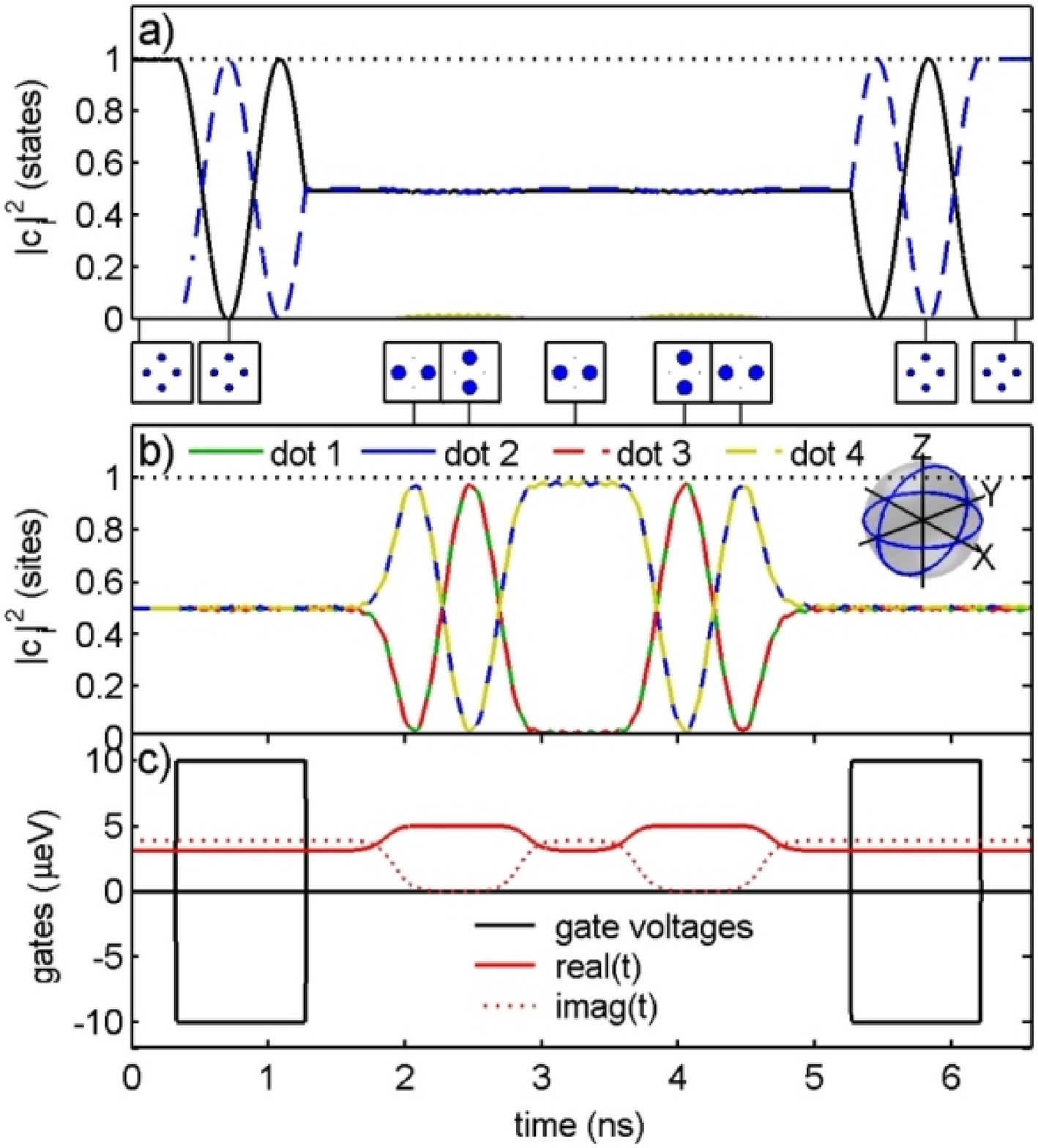}
\caption{Time evolution and control of the singlet qu2LS on the
$2\times 2$ array. (a) Time evolution of the state occupancy with
respect to the qubit basis $\left\vert 0\right\rangle _{qb}$ and
$\left\vert 1\right\rangle _{qb}$ (see
Fig.~\ref{fig_2x2_spectrum}d). (b) Time evolution of the site
occupancy $\left\vert \left\langle c_{i}^{+}c_{i}\right\rangle
\right\vert ^{2}\equiv \left\vert \left\langle \protect\psi
\left\vert c_{i}^{+}c_{i}\right\vert \protect\psi \right\rangle
\right\vert ^{2}$. The square panels in between panels (a) and (b)
show snapshots of the charge distribution on the array at the
times indicated either towards panel (a) or panel (b). The inset
in panel (b) shows the time evolution of the qubit in Bloch sphere
representation. (c) Time dependence of the voltage gates (black)
and the magnetic field expressed through $\mathrm{{Re}\left(
t\right)}$ and $\mathrm{{Im}\left( t\right)}$ (red lines) where
$\mathrm{{Abs}\left(t\right)}$ is kept constant. The time constant
for rise and fall time of the gate voltages was chosen as
$\protect\tau _{V}\equiv 0.658~\mathrm{ps}$ while for the
tunneling the considerably longer $\protect\tau _{\protect\varphi
}\equiv 100\cdot \protect\tau _{V}=65.8~\mathrm{ps}$ was used out
of adiabatic purposes with respect to the higher lying states.}
\label{fig_2x2_RK}
\end{figure}

The time evolution of the $2\times 2$ qudot array is studied
numerically for the singlet state under the action of the $V$-gate
($\sigma _z$) and the $B$-gate ($\sigma _x$). Note that here the
$B$-gate is considered non-active when a magnetic field is tuned
to the phase $\varphi =\varphi _{0}$ indicated by the arrow in
Fig.~\ref{fig_2x2_spectrum}b, while the gate is considered active
when the magnetic field is turned off ($\varphi =0$). In this
sense, with none of the two gates applied, the system is
\textit{static} since the singlet states are degenerate. Typical
time dynamics data is shown in Fig.~\ref{fig_2x2_RK}. Starting in
the singlet ground state (Bloch vector $\vec{r}_{b}=+\hat{x}$),
the system is static at $t=0$. The following $V$-gate rotates this
state around the $\hat{z}$ axis by $450^{\circ }$, ending in
$\vec{r}_{b}=+\hat{y}$, where the system is stalled for a short
time interval until the $B$-gate is activated at $t=1.6\unit{ns}$.
Now the $B$-gate rotates $\vec{r}_{b}$ around the $\hat{x}$-axis
again by $450^{\circ }$, leaving the state in the $+\hat{z}$ state
where the system is stalled again at $t=3\unit{ns}$, now being in
the central region on the time axis in figure
Fig.~\ref{fig_2x2_RK}b. Since $\pm \hat{z}$ corresponds to the
basis states of the qubit, the charge distribution equals the
$\left\vert 1\right\rangle _{qb}$ state in
Fig.~\ref{fig_2x2_spectrum}d as can be seen by the snapshots shown
along the time evolution in between panels (a) and (b) in
Fig.~\ref{fig_2x2_RK}. After another $B$-gate of the same
duration, the system is rotated another $450^{\circ }$ around the
$\hat{x}$-axis leaving the qubit in the $-\hat{y}$ state at
$t=4.8\unit{ns}$. When finally a $(-V)$-gate is applied, the qubit
is rotated by $-450^{\circ }$ around the $\hat{z}$-axis and the
system is left in the $-\hat{x} $ configuration at
$t=6.2\unit{ns}$. The time evolution of the Bloch vector in this
whole process sweeps two grand circles in the Bloch sphere, as
shown in the inset of panel (b).

The numerical data shown in Fig.~\ref{fig_2x2_RK} confirms the
previous analysis. The main point is that the qubit can be placed
into any state by applying appropriate magnetic field and gate
voltages to the system, yielding full control of the qubit as
desired.

\section{Conclusions}

Singlet and triplet states have been studied on a planar array of
quantum dots. With respect to qubits, a two-fold nearly degenerate
ground state pair was constructed and the splitting of this
two-level system was explained and estimated using the Feshbach
formalism. Furthermore these states were shown to represent full
single qubit operations encoded in the charge states. The AB flux
given by a uniform magnetic field provides the required second
quantum gate by generating a dynamical phase in the wave function
of the two electron qubit state. This can be used to effectively
turn off the qubit dynamics at will.

\begin{acknowledgements}
We acknowledge helpful discussions with A.~Govorov and with
D.~Phillips (especially for suggesting to us the Feshbach
formalism), as well as support from NSF Grant NIRT 0103034, and
the Condensed Matter and Surface Sciences Program at Ohio
University.
\end{acknowledgements}

\appendix

\section{Feshbach formalism $\label{APP.Feshbach}$}

The Feshbach formalism provides an efficient procedure to reduce
the full Hamiltonian of a large (possibly infinite) system to the
Hamiltonian of a small (finite) subsystem which should be
energetically well separated from the remainder of the space.
\cite{Feshbach62} Given a finite subspace $\mathcal{P}$ of the
total Hilbert space $\mathcal{H}$ with its complement
$\mathcal{Q}$, such that $\mathcal{P+Q=H}$, the projections into
these spaces are $P\equiv \sum_{i\in \mathcal{P} }\left\vert
i\right\rangle \left\langle i\right\vert $ and $Q\equiv
\sum_{k\notin \mathcal{P}}\left\vert k\right\rangle \left\langle
k\right\vert =1-P$, respectively. The stationary Schr\"{o}dinger
equation when projected into the $\mathcal{P}$ and $\mathcal{Q}$
spaces becomes
\begin{equation}
\begin{pmatrix}
H_{PP} & H_{PQ} \\
H_{QP} & H_{QQ}%
\end{pmatrix}%
\begin{pmatrix}
\left\vert \psi _{P}\right\rangle \\
\left\vert \psi _{Q}\right\rangle%
\end{pmatrix}%
=E%
\begin{pmatrix}
\left\vert \psi _{P}\right\rangle \\
\left\vert \psi _{Q}\right\rangle%
\end{pmatrix}
\label{fb_H}
\end{equation}%
with the projections defined as $H_{PQ}\equiv PHQ$, $\left\vert
\psi _{P}\right\rangle \equiv P\left\vert \psi \right\rangle $,
and similarly for the remaining ones. By assumption, $\mathcal{P}$
has finite dimension, thus $H_{PP}$ is also finite. $\left\vert
\psi _{Q}\right\rangle $ can be formally eliminated from
Eq.~(\ref{fb_H}) and the result is $H_{\mathit{eff}}^{P}\left\vert
\psi _{P}\right\rangle =E\left\vert \psi _{P}\right\rangle $ with
\begin{equation}
H_{\mathit{eff}}^{P}\equiv H_{PP}+\underset{\equiv \Sigma
_{QQ}\left( E\right)
}{\underbrace{H_{PQ}\frac{1}{E-H_{QQ}}H_{QP}}}, \label{fb_Heff}
\end{equation}
with $H_{\mathit{eff}}^{P}$ describing the effective Hamiltonian
in space $\mathcal{P}$. $H_{\mathit{eff}}^{P}$ is thus the sum of
the unperturbed matrix elements $H_{PP}$ and the
\textit{self-energy }contribution $\Sigma _{QQ}$. Note that
Eq.~(\ref{fb_Heff}) is still exact and thus must be nonlinear in
$E$ in order to represent the entire eigenspectrum of the system.
The non-linearity is manifested in the $E$ dependence of $\Sigma
_{QQ}$. It ensures a good approximation while it does not result
in further complications here, since for the two-dimensional space
$\mathcal{P}$ discussed in this paper, the resulting equations can
be conveniently solved analytically.

If the full Hamiltonian $H$ did not couple the $\mathcal{P}$ and
$\mathcal{Q}$ spaces, then with $H_{PQ}=0$ the second term $\Sigma
_{QQ}$ in Eq.~(\ref{fb_Heff}) vanishes consistently. Note also
that eliminating $\left\vert \psi _{Q}\right\rangle $ from
Eq.~(\ref{fb_H}) eliminates the coefficients of $\left\vert \psi
_{Q}\right\rangle $ so that this procedure systematically
eliminates variables on the large scale. Eq.~(\ref{fb_Heff}) is a
formal solution since the self-energy $\Sigma _{QQ}\left( E\right)
$ is not known. But there are different ways to approximate
$\Sigma _{QQ}$ using perturbative expansions naturally suggested
by its definition. With the identity
\begin{equation}
\frac{1}{\omega -\left( H_{0}+V\right) _{QQ}}=\frac{1}{\omega -H_{0,QQ}}%
\sum_{n=0}^{\infty }\left( V_{QQ}\frac{1}{\omega -H_{0,QQ}}\right)
^{n}, \label{fb_G_expansion}
\end{equation}%
a very useful generalization of the Brillouin-Wigner formalism to
more than one, but still a finite number of states, is obtained.
\cite{March67} This approach is essentially a path formulation in
Hilbert space in the sense that \textit{all possible path
histories through Hilbert space} are taken, starting in
$\mathcal{P}$, proceeding directly into $\mathcal{Q}$ ($H_{PQ}$ in
Eq.~\ref{fb_Heff}) and only in the final step returning back to
$\mathcal{P}$ ($H_{QP}$ in Eq.~\ref{fb_Heff}). In addition, every
intermediate step is weighted by the \textit{propagator } terms
$\left\langle k\right\vert \left( E-H_{QQ}^{0}\right)
^{-1}\left\vert k\right\rangle =\left( E-\varepsilon _{k}\right)
^{-1}$ (Eq.~\ref{fb_G_expansion}). The lowest order contributions
are then equivalent to the shortest histories through
$\mathcal{Q}$ space with lowest cost in energy.

\end{document}